\documentclass[]{article}
\usepackage{amsfonts}
\usepackage{mathtools}
\usepackage{amsmath}
\usepackage{amsthm,amssymb}
\usepackage{centernot}
\usepackage{color,soul}
\usepackage{xcolor}
\usepackage{soul}

\usepackage{graphicx}
\usepackage{subcaption}
\usepackage{epstopdf}
\usepackage{enumitem} 
 
\newtheorem{theorem}{Theorem}
\newtheorem{proposition}{Proposition}
\newtheorem{lemma}{Lemma}

\newtheorem{remark}{Remark}
\newtheorem{assumption}{Assumption}

\newcommand{\beq}{\begin{equation}}
\newcommand{\eeq}{\end{equation}}
\newcommand{\beqa}{\begin{eqnarray}}
\newcommand{\eeqa}{\end{eqnarray}}
\newcommand {\dd}   {{\rm d}}

\newcommand {\e} {\varepsilon}

\title{Evolutionary dynamics in an SI epidemic model with phenotype-structured susceptible compartment}

\author{Tommaso Lorenzi\footnote{Department of Mathematical Sciences ``G. L. Lagrange'', Dipartimento di Eccellenza 2018-2022, Politecnico di Torino, 10129 Torino, Italy. E-mail: tommaso.lorenzi@polito.it}, Andrea Pugliese\footnote{Department of Mathematics, Universit\`a di Trento. Via Sommarive, 14 - 38123 Povo (TN), Italy. E-mail: andrea.pugliese@unitn.it}, Mattia Sensi\footnote{Department of Mathematics, Universit\`a di Trento. Via Sommarive, 14 - 38123 Povo (TN), Italy. E-mail: mattia.sensi@unitn.it}, Agnese Zardini\footnote{Department of Mathematics, Universit\`a di Trento. Via Sommarive, 14 - 38123 Povo (TN), Italy. E-mail: agnese.zardini@unitn.it}}

\date{}

\begin{document}

\maketitle
\vspace{-0.8cm}
\begin{abstract}
We present an SI epidemic model whereby a continuous structuring variable captures variability in proliferative potential and resistance to infection among susceptible individuals. The occurrence of heritable, spontaneous changes in these phenotypic characteristics and the presence of a fitness trade-off between resistance to infection and proliferative potential are explicitly incorporated into the model. The model comprises an ordinary differential equation for the number of infected individuals that is coupled with a partial integrodifferential equation for the population density function of susceptible individuals through an integral term. The expression for the basic reproduction number $\mathcal{R}_0$ is derived, the disease-free equilibrium and endemic equilibrium of the model are characterised and a threshold theorem involving $\mathcal{R}_0$ is proved. Analytical results are integrated with the results of numerical simulations of a calibrated version of the model based on the results of artificial selection experiments in a host-parasite system. The results of our mathematical study disentangle the impact of different evolutionary parameters on the spread of infectious diseases and the consequent phenotypic adaption of susceptible individuals. In particular, these results provide a theoretical basis for the observation that infectious diseases exerting stronger selective pressures on susceptible individuals and being characterised by higher infection rates are more likely to spread. Moreover, our results indicate that heritable, spontaneous phenotypic changes in proliferative potential and resistance to infection can either promote or prevent the spread of infectious diseases depending on the strength of selection acting on susceptible individuals prior to infection. Finally, we demonstrate that, when an endemic equilibrium is established, higher levels of resistance to infection and lower degrees of phenotypic heterogeneity among susceptible individuals are to be expected in the presence of infections which are characterised by lower rates of death and exert stronger selective pressures. 
\end{abstract}

\section{Introduction}
Mathematical models have been widely used in epidemiology to gain a deeper understanding of infectious diseases, make predictions of disease dynamics and inform possible strategies for controlling disease outbreaks~\cite{brauer2017mathematical,grassly2008mathematical,kermack1927contribution,thompson2019detection}. 

A well-established mathematical approach to describing the evolution of infectious diseases is to use compartment models formulated in terms of systems of ordinary differential equations (ODEs), whereby each differential equation governs the dynamic of the size of a compartment~\cite{2,3,iannelli2015introduction}. This approach traditionally relies on the assumption that phenotypic characteristics are homogeneously distributed among individuals in the same compartment (e.g. susceptible individuals are all characterised by the same disease susceptibility, and all infected individuals have the same death and recovery rates). However, such a simplifying assumption is only prima facie justified and may narrow down the application domain of these epidemic models considerably.  For instance, a mounting body of empirical evidence shows that there are situations in which the evolution of infectious diseases may be importantly shaped by phenotypic heterogeneity among susceptible individuals~\cite{chabas2018evolutionary,de2004host,gomes2012host,marm2006host,osnas2012evolution,Pugliese2011a,stadler2013uncovering}. 

A possible way of incorporating phenotypic heterogeneity in compartment models consists in structuring one or more compartments by a continuous variable that provides a mathematical representation of the phenotype of the individuals (i.e. individuals with different phenotypes are characterised by different values of the structuring variable). The phenotypic distribution of individuals in a structured compartment is described by a population density function, the dynamic of which is governed by an integrodifferential equation (IDE) or a partial integrodifferential equation (PIDE) that replaces the ODE for the size of the corresponding unstructured compartment in the original model. Moreover, some of the parameters of the unstructured counterpart of the model are replaced by suitable functions of the structuring variable and the model equations may be coupled through integral terms.

Epidemic models that comprise structured compartments have been proposed in many different contexts, from age-structured populations~\cite{busenberg1991global,inaba1990threshold,inaba2017age,thieme2009spectral} to populations occupying spatially heterogeneous environments \cite{inaba2012new,lou2011reaction,peng2012reaction,wang2011nonlocal,wang2012basic}. Closer to the topic of our study, we mention some papers in which susceptibility and/or infectivity depend on a continuous structuring variable~\cite{D,A,F,E,C}, leading to an IDE for the population density of the corresponding compartment.

We focus here on heterogeneity in susceptibility to infection, which we assume to depend on a structuring variable that regulates also the proliferative potential of the individuals, giving rise to a trade-off between resistance to infection and reproduction rate. Such a fitness trade-off has been experimentally shown to play a key role in the adaptation of various species  to the selective pressure exerted by different forms of infection. For instance, in~\cite{gustafsson1994infectious} it was reported that collared flycatcher (\textit{Ficedula albicollis}) exhibits a trade-off between reproduction rate and resistance to parasites. The same kind of trade-off was documented and studied in~\cite{boots1999evolution}, using a more theoretical approach. In \cite{festa1989individual}, the author found a positive correlation, in a population of bighorn ewes (\textit{Ouis Canadensis}), between resources allocated for reproduction and the presence of a specific parasite. In~\cite{lochmiller2000trade} and~\cite{sheldon1996ecological}, the authors report on various other examples of species in which this trade-off is present. In~\cite{WeWo}, the existence of a fitness trade-off between resistance to a parasite (\textit{Schistosoma mansoni}) and fertility was observed in a species of freshwater snail (\textit{Biomphalaria glabrata}).

The variable controlling reproduction rate and resistance to infection is subjected to evolutionary pressures that vary depending on the number of infective individuals. We further assume the occurrence of heritable, spontaneous changes in these phenotypic characteristics. Phenotypic changes are incorporated into our model via a diffusion term~\cite{genieys2006adaptive}.

Models coupling epidemic processes with evolutionary changes  have been studied by Burie, Ducrot and co-workers in a series of papers~\cite{abi2021asymptotic,burie2020asymptotic,burie2020slow,burie2020concentration,djidjou2017steady}. For instance, in~\cite{djidjou2017steady}, the authors considered a system with three compartments, that is, a homogeneous compartment of susceptible hosts, a heterogeneous compartment of infective agents (pathogen spores) and an age-structured compartment of infected hosts. A model with a homogeneous susceptible compartment and a continuously-structured infective compartment was also considered in~\cite{day2004general}; the authors then reduced the model to a system of ordinary differential equations, one dependent variable of which represented the mean virulence.

To the best of our knowledge, there are no published papers that consider a model that includes a continuous phenotypic structure in the \emph{susceptible} compartment and allows for changes in this phenotypic structure to occur, under the evolutionary pressure associated with a fitness trade-off between reproductive potential and susceptibility. This enables a mathematical dissection of the roles played by different evolutionary parameters in the phenotypic adaptation of the host population to the selective pressure exerted by the infection.

We frame our analysis in the context of an SI epidemic model, as it is common in evolutionary epidemiology~\cite{AndMay81,ResKoe04}. The model comprises an ODE for the number of infected individuals which is coupled with a PIDE for the population density function of susceptible individuals through an integral term. Exploiting the analytical tractability of the model equations, we obtain a detailed mathematical depiction of the interactions between susceptible and infected individuals. In summary, the expression for the basic reproduction number $\mathcal{R}_0$ is derived, the  disease-free equilibrium and endemic equilibrium of the model are characterised and a threshold theorem involving $\mathcal{R}_0$ is proved. Analytical results are integrated with the results of numerical simulations of a calibrated version of the model based on the results of artificial selection experiments presented in~\cite{WeWo}. The analytical and numerical results obtained clarify the role of different evolutionary parameters on the spread of infectious diseases and the consequent phenotypic adaption of susceptible individuals. 

The paper is organised as follows. In Section~\ref{Model} we describe the mathematical model. In Section~\ref{Analytics} we present the main analytical results of our study. In Section~\ref{Numerics} we integrate analytical results with numerical simulations, and we discuss the biological relevance of our theoretical findings. Section~\ref{Conclusions} concludes the paper and provides a brief overview of possible research perspectives.     

\section{Description of the model}
\label{Model}
\paragraph{Model equations.} We consider an SI epidemic model whereby the susceptible compartment is structured by a continuous variable $x \in X$, with $X$ being an interval (possibly infinite) of $ \mathbb{R}$, which models the phenotypic state of every susceptible individual and takes into account interindividual variability in proliferative potential (i.e. the number of progeny produced per unit time) and resistance to infection. The phenotype distribution of susceptible individuals at time $t \geq 0$ is described by the population density function $s(x,t) \geq 0$, while the number of infected individuals at time $t$ is modelled by the function $I(t) \geq 0$. The size of the susceptible compartment (i.e. the number of susceptible individuals) and the total number of individuals are defined, respectively, as
\beq
\label{eq:S}
S(t) := \int_{X} s(x,t) \, \dd x \quad \text{and} \quad N(t) := S(t) + I(t).
\eeq
Moreover, the mean phenotypic state of susceptible individuals and the related variance are defined, respectively, as 
\beq
\label{eq:XV}
\mu(t) := \frac{1}{S(t)} \, \int_{X} x\, s(x,t) \, \dd x \quad \text{and} \quad \sigma^2(t) := \frac{1}{S(t)} \, \int_{X} x^2\, s(x,t) \, \dd x - \mu(t)^2. 
\eeq
The function $\sigma^2(t)$ provides a measure of the degree of phenotypic heterogeneity in the susceptible compartment. 

The evolution of the functions $s(x,t)$ and $I(t)$ is governed by the following PIDE-ODE system
\beq\label{eq:syst}
\begin{cases}
\displaystyle{\dfrac{\partial s}{\partial t} = \left(a(x) - \frac{N(t)}{K}\right) s - b(x) \, I \, s  + \beta \dfrac{\partial^2 s}{\partial x^2}}, \quad (x,t) \in X \times (0,\infty),
\\\\
\displaystyle{I' =  \left(\int_{X} b(x) \, s(x,t) \, \dd x  - \nu \right) I,} \quad\quad\quad\quad\quad\quad t \in (0,\infty),
\\\\
\displaystyle{N(t) := S(t) + I(t), \quad S(t) := \int_{X} s(x,t) \, \dd x,}
\end{cases}
\eeq
subject to a biologically relevant initial condition of components
\beqa
\label{eq:ICgen}
& s(x,0) = s^0 \in C(\mathbb{R}), \quad s^0(\cdot) \geq 0, \quad 0 < S(0) := \displaystyle{\int_{X} s^0(x) \, \dd x} < \infty, \nonumber 
\\\\
& I(0) = I^0 > 0 \nonumber
\eeqa
and to zero Neumann (i.e. no-flux) boundary conditions at the endpoints of $X$ in the case where the interval $X$ is bounded.

In the PIDE~\eqref{eq:syst}$_1$, the diffusion term models the effect of heritable, spontaneous phenotypic changes, which occur at rate $\beta>0$. The function $a(x)$ is the  net per capita growth rate of susceptible individuals in the phenotypic state $x$ in the absence of infected individuals (i.e. the intrinsic net per capita growth rate), while the parameter $K>0$ is related to the carrying capacity of the system (i.e. the maximal number of individuals that can be accommodated within the system due to space limitations). Furthermore, the function $b(x)$ is the rate of infection (i.e. the product between the contact rate and the relative susceptibility of individuals in the phenotypic state $x$). Finally, the parameter $\nu > 0$ in the ODE~\eqref{eq:syst}$_2$ is the rate of death caused by the infection. Notice that we neglect physiological death of infected individuals since we assume its rate, which might depend on $N(t)$, to be much lower than the rate of death caused by the infection. Notice also that infected individuals do not reproduce since we consider the case where proliferation of infected individuals is impaired by the infection.

\paragraph{Assumptions on the functions $a(x)$ and $b(x)$.} We focus on a biological scenario whereby there is a fitness trade-off between resistance to infection and proliferative potential~\cite{dallas2016costs,Kliot,rivero2011energetic,WeWo,woolhouse1989effect}. Under this scenario, without loss of generality, we let $X$ be such that $[0,1] \subset X$ and make the following biological assumptions.

\begin{assumption}
\label{bioass1}
Susceptible individuals in the phenotypic state $x=0$ are characterised by the highest proliferative potential, whilst susceptible individuals in the phenotypic state $x=1$ have the highest resistance to infection (cf. the schematics displayed in Figure~\ref{fig0}). 
\end{assumption}

\begin{assumption}
\label{bioass3}
Natural selection is moderately intense in the absence of infection. As a result, individuals in the phenotypic state $x=1$ have a positive intrinsic net per capita growth rate. Moreover, due to natural selection, phenotypic variants that are not proliferative enough nor sufficiently resistant to infection (i.e. individuals in phenotypic states $x$ sufficiently smaller than $0$ or sufficiently larger than $1$) will have a negative intrinsic net per capita growth rate.
\end{assumption}

\begin{figure}[h!]
	\begin{center}
		\includegraphics[width=0.8\linewidth]{Fig0.png}
		\caption{Schematics illustrating the relationships between the phenotypic state $x$, the proliferative potential and the level of resistance to infection of susceptible individuals.}
		\label{fig0}
	\end{center}
\end{figure}

Considering a biological scenario corresponding to Assumptions~\ref{bioass1} and \ref{bioass3}, we let the functions $a(x)$ and $b(x)$ be such that
\beq\label{ass:agen}
a : X  \to \mathbb{R}, \;\; a \in C(X), \;\;  \max_{x \in X}  a(x) = a(0) > a(1) > 0, \;\; \lim_{|x| \to \infty} a(x) = - \infty
\eeq
and
\beq\label{ass:bgen}
b : X  \to \mathbb{R}^+_*, \quad b \in C(X), \quad  \min_{x \in X}  \, b(x) = b(1) > 0,
\eeq
where $\mathbb{R}^+_*$ denotes the set of positive real numbers.

\paragraph{Specific definitions of the functions $a(x)$ and $b(x)$.} Under assumptions~\eqref{ass:agen} and \eqref{ass:bgen}, building upon the modelling strategies presented in~\cite{Ardassseva2019,chisholm2016evolutionary,lorenzi2015dissecting}, we will focus on the case where
\beq\label{eq:ab}
 a(x) := \gamma - \eta \, x^2 \quad \text{with} \quad \gamma>\eta \quad \text{and} \quad b(x) := \xi +  \zeta \, (1-x)^2.
 \eeq
In definitions~\eqref{eq:ab}, the parameter $\gamma > 0$ models the intrinsic net per capita growth rate of susceptible individuals in the highly proliferative phenotypic state $x=0$ (i.e. $\gamma$ is the maximum intrinsic net per capita growth rate), and the parameter $\eta>0$ is a selection gradient that quantifies the selective pressure acting on susceptible individuals prior to infection (i.e. $\eta$ is the intrinsic selection gradient). The parameter $\xi > 0$ is the rate at which the infection is transmitted from infected individuals to susceptible individuals in the highly resistant phenotypic state $x=1$ (i.e. $\xi$ is the minimal infection rate), and the parameter $\zeta>0$ is a selection gradient that quantifies the intensity of the selective pressure that the infection exerts on susceptible individuals (i.e. $\zeta$ is the selection gradient related to infection). 

\begin{remark}
\label{remeffecselec}
If the functions $a(x)$ and $b(x)$ are defined according to~\eqref{eq:ab}, a little algebra shows that the difference $a(x) - b(x) \, I$ on the right-hand side of the PIDE~\eqref{eq:syst}$_1$ can be rewritten as
\begin{equation}\label{pn2ori}
a(x) - b(x) \, I = g(I) - h(I) \left(x - \varphi(I) \right)^2
\end{equation}
where
\begin{equation}\label{ghori}
g(I) := \gamma - \left(\frac{\eta \, \zeta}{\eta + \zeta I} + \xi \right) I, \quad h(I) :=  \eta + \zeta I, \quad \varphi(I) := \frac{\zeta I}{\eta + \zeta I}.
\end{equation}
From the point of view of evolutionary dynamics, the function $g(I)$ can be seen as the maximum fitness of susceptible individuals under the environmental conditions corresponding to the number of infected individuals $I$. Moreover, the function $h(I)$ is an effective selection gradient that quantifies the total selective pressure exerted on susceptible individuals in the phenotypic state $x$ when $I$ infected individuals are present. Finally, the function $\varphi(I)$ is the fittest phenotypic state in the presence of $I$ infected individuals. Notice that, consistent with our modelling assumptions (see the schematics displayed in Figure~\ref{fig0}), we have
$$
\varphi : \mathbb{R}^+ \to [0,1] \quad \text{and} \quad \lim_{I\to0} \varphi(I) = 0, \quad \lim_{I\to\infty} \varphi(I) = 1.
$$  
\end{remark}

\section{Analytical results}
\label{Analytics}
In this section, we characterise the disease-free equilibrium of the model and derive the expression for the basic reproduction number $\mathcal{R}_0$ (see Section~\ref{sec:an1}). Furthermore, we characterise the endemic equilibrium of the model (see Section~\ref{sec:an2}). Finally, we study the long-time asymptotic behaviour of the model equations by proving a threshold theorem involving $\mathcal{R}_0$ (see Section~\ref{sec:an3}).

\subsection{Disease-free equilibrium and basic reproduction number}
\label{sec:an1}
\paragraph{Disease-free equilibrium.} The population density function of susceptible individuals at disease-free equilibrium $\overline{s}_F(x)$ satisfies the following steady-state problem 
\beq\label{eq:systssF}
\begin{cases}
\displaystyle{\beta \, \overline{s}_F'' + \left(a(x) - \frac{\overline{S}_F}{K} \right)\, \overline{s}_F = 0, \quad x \in X}, 
\\\\
\displaystyle{\overline{S}_F := \int_{X} \overline{s}_F(x) \, \dd x},
\\\\
\overline{s}_F(\cdot) > 0, \quad  0 < \overline{S}_F < \infty,
\end{cases}
\eeq
subject to zero Neumann boundary conditions if $X$ is bounded.

In the case where the intrinsic net per capita growth rate $a(x)$ satisfies assumptions~\eqref{ass:agen}, a characterisation of $\overline{s}_F(x)$ is provided by Theorem~\ref{th:zero}, the proof of which relies on the facts recalled in Lemma~\ref{lem:zero} about the principal eigenpair $(\lambda,\psi)$ of the elliptic differential operator 
\beq
\label{def:L}
\mathcal{L} := \beta \Delta + a
\eeq
(i.e. $\mathcal{L}[\psi] = - \lambda \psi$) acting on functions defined on $X$. These are well known facts in the case where $X$  is a bounded interval and Neumann boundary conditions are considered. The case where $X \equiv \mathbb{R}$ is considered, for instance, in~\cite[Theorem 2.3.1]{berezinshubin}, while possible generalisations can be found in~\cite{berestycki1994principal,berestycki2015generalizations}. Notice that the characterisation of the principal eigenvalue given by~\eqref{def:rayleigh} follows immediately, once the existence of a complete set of discrete eigenfunctions of $\mathcal{L}$ has been shown.

\begin{lemma}
\label{lem:zero}
When $a(x)$ satisfies assumptions~\eqref{ass:agen}, the principal eigenvalue $\lambda$ of the elliptic differential operator $\mathcal{L}$ is simple and there is a unique normalised positive eigenfunction $\psi$ associated with $\lambda$. The eigenfunction $\psi$ is smooth and the principal eigenvalue $\lambda$ is given by
$$
\lambda = \inf_{\phi \in{H^1(X) \setminus \{0\}}} \mathcal{Q}(\mathcal{L}, \phi),
$$
with 
\beq
\label{def:rayleigh}
 \mathcal{Q}(\mathcal{L},\phi) := \frac{\beta \displaystyle{\int_X \left|\nabla \phi(x) \right|^2 \, {\rm d}x - \int_X a(x) \phi^2(x) \, {\rm d}x}}{\displaystyle{\int_X \phi^2(x) \, {\rm d}x}}.
\eeq
\end{lemma} 

\begin{theorem} 
\label{th:zero}
Let assumptions~\eqref{ass:agen} hold and assume $a(x)$ and $\beta$ to be such that
\beq\label{ass:noextgen}
\inf_{\phi \in{H^1(X) \setminus \{0\}}} \mathcal{Q}(\mathcal{L}, \phi) < 0.
\eeq
Then, there exists a unique solution of problem~\eqref{eq:systssF}, subject to zero Neumann boundary conditions if $X$ is bounded, which is given by 
\beq
\label{eq:sfgen}
\overline{s}_F(x) = \overline{S}_F \, \psi(x) \quad \text{with} \quad \overline{S}_F = K \, \int_{X} a(x) \, \psi(x) \, \dd x.
\eeq
\end{theorem} 

\begin{proof}
The differential equation~\eqref{eq:systssF}$_1$ can be rewritten as  
$$
\mathcal{L}[\overline{s}_F] =  \frac{\overline{S}_F}{K} \, \overline{s}_F, \quad x \in X,
$$
that is, $-\dfrac{\overline{S}_F}{K}$ is an eigenvalue of the elliptic differential operator $\mathcal{L}$. Hence, Lemma~\ref{lem:zero} allows us to conclude that, under assumptions~\eqref{ass:agen} and~\eqref{ass:noextgen}, there exists a unique solution of problem~\eqref{eq:systssF}, subject to zero Neumann boundary conditions if $X$ is bounded, which is given by
\beq
\label{sFtemp}
\overline{s}_F(x) = \overline{S}_F \, \psi(x) \quad \text{with} \quad \overline{S}_F = - K \, \lambda > 0.
\eeq
Furthermore, integrating both sides of the differential equation~\eqref{eq:systssF}$_1$ over $X$, dividing through by $\overline{S}_F$ and substituting the expression~\eqref{sFtemp} for $\overline{s}_F(x)$ into the resulting equation gives
$$
\overline{S}_F = K \, \int_{X} a(x) \, \psi(x) \, \dd x.
$$
This concludes the proof of Theorem~\ref{th:zero}.
\end{proof}

If $X \equiv \mathbb{R}$ and the intrinsic net per capita growth rate $a(x)$ is defined according to~\eqref{eq:ab}, the population density function~$\overline{s}_F(x)$ can be explicitly characterised, as shown by Proposition~\ref{prop:zero}. The proof of Proposition~\ref{prop:zero} is similar to that of Lemma~3.2 in~\cite{chisholm2016evolutionary} and, therefore, it is omitted here.
\begin{proposition} 
\label{prop:zero}
Let $X \equiv \mathbb{R}$. If the function $a(x)$ is defined via~\eqref{eq:ab} and 
\beq
\label{ass:noextspec}
\gamma > \left(\eta \beta \right)^{\frac{1}{2}},
\eeq
then the unique solution of problem~\eqref{eq:systssF} is
\beq
\label{eq:st1}
\overline{s}_F(x) = \overline{S}_F \ \left(\frac{1}{2 \pi \, \overline{\sigma}^2_F}\right)^{\frac{1}{2}} \exp\left[-\frac{(x-\overline{\mu}_F)^2}{2 \, \overline{\sigma}^2_F} \right],
\eeq
with
\beq
\label{eq:st1barS}
\overline{\sigma}^2_F = \left(\frac{\beta}{\eta}\right)^{\frac{1}{2}}, \quad \overline{\mu}_F= 0, \quad \overline{S}_F = K \left(\gamma - \left(\eta \, \beta \right)^{\frac{1}{2}}\right) > 0.
\eeq
\end{proposition} 
 
\paragraph{Basic reproduction number $\mathcal{R}_0$.} In the modelling framework of~\eqref{eq:syst}, the basic reproduction number $\mathcal{R}_0$ is defined as
\beq\label{defR0gen}
\mathcal{R}_0 := \dfrac{1}{\nu} \int_{X} b(x) \, \overline{s}_F(x) \, \dd x.
\eeq
Under assumptions~\eqref{ass:agen} and~\eqref{ass:noextgen}, substituting the expression~\eqref{eq:sfgen} for the population density function of susceptible individuals at disease-free equilibrium into~\eqref{defR0gen} gives
\beq
\label{defR0a}
\mathcal{R}_0 = \frac{\overline{S}_F}{\nu} \int_{X} b(x) \, \psi(x) \, \dd x \quad \text{with} \quad \overline{S}_F = K \, \int_{X} a(x) \, \psi(x) \, \dd x.
\eeq
In particular, when the intrinsic net per capita growth rate $a(x)$ and the infection rate $b(x)$ are defined via~\eqref{eq:ab} and assumption~\eqref{ass:noextspec} holds, substituting the expression~\eqref{eq:st1} for the population density function~$\overline{s}_F(x)$ and definition~\eqref{eq:ab} of the infection rate $b(x)$ into~\eqref{defR0gen}, with a little algebra one finds the following explicit expression for the basic reproduction number
\beq
\label{defR0b}
\mathcal{R}_0 = \frac{K}{\nu} \left(\gamma -(\eta \beta)^{\frac{1}{2}}\right) \left(\xi +\zeta+ \zeta  \left(\dfrac{\beta}{\eta}\right)^{\frac{1}{2}}\right).
\eeq

\subsection{Endemic equilibrium}
\label{sec:an2}
The population density function of susceptible individuals at endemic equilibrium $\overline{s}_E(x)$ and the corresponding number of infected individuals $\overline{I}_E$ satisfy the following steady-state problem 
\beq\label{eq:systssE}
\begin{cases}
\displaystyle{\beta \, \overline{s}_E'' + \left[a(x) - \left(b(x) + \frac{1}{K}\right) \, \overline{I}_E  - \frac{\overline{S}_E}{K} \right]\, \overline{s}_E = 0, \quad x \in X}, 
\\\\
\displaystyle{\left(\int_{\mathbb{R}} b(x) \, \overline{s}_E(x) \, \dd x  - \nu \right) \, \overline{I}_E = 0},
\\\\
\displaystyle{\overline{S}_E := \int_{X} \overline{s}_E(x) \, \dd x},
\\\\
\overline{s}_E(\cdot) > 0, \quad  0 < \overline{S}_E < \infty, \quad  0 < \overline{I}_E < \infty,
\end{cases}
\eeq
subject to zero Neumann boundary conditions if $X$ is bounded.

In the case where the intrinsic net per capita growth rate $a(x)$ and the infection rate $b(x)$ satisfy assumptions~\eqref{ass:agen} and~\eqref{ass:bgen} and the additional assumption~\eqref{ass:noextgen} holds, a characterisation of $\overline{s}_E(x)$ and $\overline{I}_E$ is provided by Theorem~\ref{th:uno}. The proof of Theorem~\ref{th:uno} relies on the well known facts recalled in Lemma~\ref{lem:uno} (corresponding to Lemma \ref{lem:zero}) about the principal eigenpair $(\lambda_y,\psi_y)$ of the elliptic differential operator 
\beq
\label{def:LE}
\mathcal{L}_y := \beta \Delta + r_y, \quad r_y(x) := a(x) -  \left(b(x) + \frac{1}{K} \right) y
\eeq
(i.e. $\mathcal{L}_y[\psi_y] = - \lambda_y \psi_y$) with zero Neumann boundary conditions (if $X$ is bounded) and acting on functions defined on $X$.

\begin{lemma}
\label{lem:uno}
When $a(x)$ and $b(x)$ satisfy assumptions~\eqref{ass:agen} and~\eqref{ass:bgen}, for any $y \geq 0$ the principal eigenvalue $\lambda_y$ of the elliptic differential operator $\mathcal{L}_y$ is simple, and there is a unique normalised positive eigenfunction $\psi_y$ associated with $\lambda_y$. The eigenfunction $\psi_y$ is smooth, and the principal eigenvalue $\lambda_y$ is given by
$$
\lambda_y = \inf_{\phi \in{H^1(X) \setminus \{0\}}} \mathcal{Q}(\mathcal{L}_y, \phi),
$$
with $\mathcal{Q}(\mathcal{L}_y, \phi)$ defined via~\eqref{def:rayleigh}, and is monotonically increasing in $y$.
\end{lemma} 
 
\begin{theorem} 
\label{th:uno}
Let assumptions~\eqref{ass:agen}, \eqref{ass:bgen} and \eqref{ass:noextgen} hold and assume $\mathcal{R}_0>1$. Then there exists at least a positive solution $(\overline{s}_E, \overline{I}_E)$ of problem~\eqref{eq:systssE}, subject to zero Neumann boundary conditions if $X$ is bounded, whereby $\overline{I}_E$ is a positive solution of the algebraic equation 
\beq
\label{SEtemp1}
- K \, \lambda_{\overline{I}_E} \, B_{\overline{I}_E} = \nu \quad \text{with} \quad B_y := \int_{X} b(x) \, \psi_y(x) \, \dd x
\eeq
and 
\beq
\label{eq:sfgenEproof}
\overline{s}_E(x) = \overline{S}_E \, \psi_{\overline{I}_E}(x) \quad \text{with} \quad \overline{S}_E = - K \, \lambda_{\overline{I}_E} > 0.
\eeq
Furthermore, if
\beq
\label{ass:monot}
\dfrac{\partial}{\partial y} (\lambda_y \, {B_y}) > 0
\eeq
then the solution $\overline{I}_E$ of the algebraic equation~\eqref{SEtemp1} is unique and the solution of problem~\eqref{eq:systssE} is unique as well.
\end{theorem} 

\begin{proof}
The differential equation~\eqref{eq:systssE}$_1$ can be rewritten as
$$
\mathcal{L}_{\overline{I}_E}[\overline{s}_E] =  \frac{\overline{S}_E}{K} \, \overline{s}_E, \quad x \in X,
$$
that is, $-\dfrac{\overline{S}_E}{K}$ is an eigenvalue of the elliptic differential operator $\mathcal{L}_{\overline{I}_E}$. \\
 Lemma~\ref{lem:uno} allows us to conclude that, under assumptions~\eqref{ass:agen}, \eqref{ass:bgen} and \eqref{ass:noextgen}, for $\overline{I}_E> 0$ sufficiently small so that
$$
\inf_{\phi \in{H^1(X) \setminus \{0\}}} \mathcal{Q}(\mathcal{L}_{\overline{I}_E}, \phi) < 0,
$$ 
there exists a unique solution of~\eqref{eq:systssE}$_1$ complemented with~\eqref{eq:systssE}$_3$, subject to~\eqref{eq:systssE}$_4$, and to zero Neumann boundary conditions if $X$ is bounded, which is given by 
\beq
\label{eq:sfgenEproof2}
\overline{s}_E(x) = \overline{S}_E \, \psi_{\overline{I}_E}(x) \quad \text{with} \quad \overline{S}_E = - K \, \lambda_{\overline{I}_E} > 0.
\eeq
Since $\overline{S}_E = - K \, \lambda_{\overline{I}_E}$, using the fact that $\lambda_{\overline{I}_E}$ is monotonically increasing in $\overline{I}_E$, as established by Lemma~\ref{lem:uno}, we conclude that $\overline{S}_E$ is monotonically decreasing in $\overline{I}_E$.

Substituting expression~\eqref{eq:sfgenEproof} for $\overline{s}_E(x)$ into~\eqref{eq:systssE}$_2$ gives
\beq
\label{SEtemp2}
\overline{S}_E \, B_{\overline{I}_E} = \nu
\eeq
with $B_{\overline{I}_E}$ defined via~\eqref{SEtemp1}. Since $\overline{S}_E = -  K \, \lambda_{\overline{I}_E}$, we can introduce the function
$$
F(y) := - K \, \lambda_{y} \, B_{y}
$$
and rewrite~\eqref{SEtemp2} as
$$
F( \overline{I}_E ) = \nu.
$$
Since $\mathcal{L}_0$ is the same as $\mathcal{L}$ defined in \eqref{def:L}, using \eqref{sFtemp} we find that
$$ 
F(0) = \overline{S}_F \int_{X} b(x) \, \psi(x) \, \dd x = \int_{X} b(x) \, \overline{s}_F(x) \, \dd x = \nu \, \mathcal{R}_0.
$$
Hence, under the assumption $\mathcal{R}_0 > 1$, we have that $F(0) > \nu$. On the other hand, there exists $\bar y > 0$ such that $\lambda_{\bar y} = 0$ and, therefore, $F(\bar y) = 0$. By continuity of the function $F(y)$, there exists  $\overline{I}_E \in (0, \bar y)$ such that $F( \overline{I}_E )  = \nu$.
Condition~\eqref{ass:monot} is equivalent to $F'(y) < 0$; thus, if condition~\eqref{ass:monot} is met, the solution $\overline{I}_E$ of the algebraic equation~\eqref{SEtemp1} is unique.
\end{proof}

If $X \equiv \mathbb{R}$ and the intrinsic net per capita growth rate $a(x)$ and the infection rate $b(x)$ are defined according to~\eqref{eq:ab}, the population density function~$\overline{s}_E(x)$ and the size of the infected compartment  $\overline{I}_E$ can be characterised more explicitly, as shown by Proposition~\ref{prop:uno}.
\begin{proposition}
\label{prop:uno}
Let $X \equiv \mathbb{R}$. If the functions $a(x)$ and $b(x)$ are defined via~\eqref{eq:ab}, assumption~\eqref{ass:noextspec} holds and $\mathcal{R}_0>1$, then the unique solution of problem~\eqref{eq:systssE} is given by
\beq
\label{eq:equiln}
\overline{s}_E(x) = \overline{S}_E \ \left(\frac{1}{2 \pi \, \overline{\sigma}^2_E}\right)^{\frac{1}{2}} \exp\left[-\frac{(x-\overline{\mu}_E)^2}{2 \, \overline{\sigma}^2_E} \right],
\eeq
with
\beq
\label{eq:st2barS}
\overline{\sigma}^2_E = \left(\dfrac{\beta}{\eta + \zeta \overline{I}_E}\right)^{\frac{1}{2}}, \quad \overline{\mu}_E= \frac{\zeta \overline{I}_E}{\eta +  \zeta \overline{I}_E},
\eeq
\beq
\label{eq:equilnbarS}
\overline{S}_E = K \left(\gamma - \left( \frac{1}{K} +\frac{\eta \, \zeta}{\eta + \zeta \overline{I}_E} + \xi \right) \overline{I}_E - \beta^{\frac{1}{2}} \, \left(\eta + \zeta \overline{I}_E\right)^{\frac{1}{2}}\right) > 0,
\eeq
and by $\overline{I}_E$ that is the unique positive solution of the algebraic equation
\beq
\label{eq:equilI}
\left(1+\xi K\right)\overline{I}_E = \, K\left(\gamma - \eta \frac{\zeta \overline{I}_E}{\eta + \zeta \overline{I}_E} - \beta^{\frac{1}{2}} \left(\eta + \zeta \overline{I}_E\right)^{\frac{1}{2}}\right) - F(\overline{I}_E),
\eeq
with
\begin{equation}\label{eq:AI}
F(y) := \nu \, \left(\xi +\zeta\left(\frac{\zeta y}{\eta + \zeta y}-1\right)^2+ \zeta \left(\frac{\beta}{\eta + \zeta y}\right)^{\frac{1}{2}}\right)^{-1}.
\end{equation}
\end{proposition}

\begin{proof}
Throughout the proof we drop the subscript $E$ for convenience of notation. Building upon the method of proof presented in~\cite{chisholm2016evolutionary,stace2020discrete}, first we note that when $a(x)$ and $b(x)$ are defined according to~\eqref{eq:ab} the differential equation~\eqref{eq:systssE}$_1$ can be rewritten as
\beq
\label{eqsss}
\displaystyle{\beta \, \overline{s} \, ''(x) + \left(g(\overline{I}) - \dfrac{\overline{I}}{K} - h(\overline{I})\left(x - \varphi(\overline{I}) \right)^2 - \frac{\overline{S}}{K} \right) \overline{s}(x)} = 0, \quad x \in \mathbb{R}
\eeq
where the functions $g$, $h$ and $\varphi$ are defined via~\eqref{ghori}. Then we make the change of variables 
$$
y = x - \varphi(\overline{I})
$$
and rewrite the differential equation~\eqref{eqsss} as
\begin{equation}\label{eq:ni}
\displaystyle{\beta \, \overline{s} \, ''(y) + \left(g(\overline{I}) - \dfrac{\overline{I}}{K} - h(\overline{I}) \, y^2 - \frac{\overline{S}}{K} \right) \overline{s}(y)} = 0, \quad y \in \mathbb{R}.
\end{equation} 
Finally, we make the additional change of variables 
\begin{equation}\label{eq:uchange}
\overline{s}(y) = u(z) \quad \mbox{with} \quad z = y\left(\frac{4 \, h(\overline{I})}{\beta}\right)^{\frac{1}{4}}
\end{equation} 
and in so doing we find that $u(z)$ satisfies the differential equation
\begin{equation}\label{eq:ui}
u''(z)-\left(\frac{z^2}{4} + P\right) u(z)=0, \quad z \in \mathbb{R}
\end{equation}
with
$$
P  := \frac{1}{2 \, K \left(\beta \, h(\overline{I})\right)^{\frac{1}{2}}}\left(\overline{S} - K \, g(\overline{I}) + \overline{I}\right).
$$ 
The differential equation~\eqref{eq:ui} is the Weber's equation, the solutions of which are bounded and non-negative for all $z \in \mathbb{R}$ if and only if $P = - \dfrac{1}{2}$, i.e. if and only if 
\begin{equation}\label{eq:barrho}
\overline{S} = K \left(g(\overline{I}) -  \dfrac{\overline{I}}{K} - \left(\beta \, h(\overline{I})\right)^{\frac{1}{2}} \right).
\end{equation}
From the expression~\eqref{eq:barrho} for $\overline{S}$ we see that if assumption~\eqref{ass:noextspec} holds then for $\overline{I}$ small enough we have $\overline{S}>0$.

The bounded and non-negative solutions of the differential equation~\eqref{eq:ui} are of the form
\begin{equation}\label{eq:uisol}
u(z) \propto \exp\left(-\frac{z^2}{4}\right) \,
.
\end{equation}
Combining \eqref{eq:uisol} and~\eqref{eq:uchange} yields
$$
\overline{s}(x) = c \exp \left[-\frac{1}{2}\left(\frac{h(\overline{I})}{\beta}\right)^{\frac{1}{2}}\left(x - \varphi(\overline{I})\right)^2\right]
$$
for some real constant $c$. Moreover, using~\eqref{eq:systssE}$_3$, we evaluate the constant $c$ in terms of $\overline{S}$ as
$$
c = \frac{\overline{S}}{(2\pi)^{\frac{1}{2}}} \, \left(\frac{h(\overline{I})}{\beta} \right)^{\frac{1}{4}}
$$
and conclude that
\begin{equation}\label{eq:nisol}
 \overline{s}(x) = \frac{\overline{S}}{(2\pi)^{\frac{1}{2}}} \left(\frac{h(\overline{I})}{\beta}\right)^{\frac{1}{4}}\exp\left[-\frac{1}{2}\left(\frac{h(\overline{I})}{\beta}\right)^{\frac{1}{2}}\left(x - \varphi(\overline{I})\right)^2\right].
\end{equation}
If $\overline{I}$ and $\overline{s}(x)$ satisfy conditions~\eqref{eq:systssE}$_4$ then equation~\eqref{eq:systssE}$_2$ gives
\beq\label{nr2}
\int_{\mathbb{R}} b(x) \, \overline{s}(x) \, \dd x = \nu. 
\eeq
Substituting~\eqref{eq:nisol} into~\eqref{nr2} gives
\begin{equation}\label{eq:esse1}
\overline{S} = \frac{\nu}{\displaystyle \xi +\zeta\left(\varphi(\overline{I})-1\right)^2+\zeta \bigg(\frac{\beta}{h(\overline{I})}\bigg)^{\frac{1}{2}}} =: F({\overline{I}}).
\end{equation}
Equating expressions~\eqref{eq:barrho} and \eqref{eq:esse1} for $\overline{S}$ and rearranging terms we find
\begin{equation}\label{eq:exi}
\left(1+\xi K\right)\overline{I} = K\left(\gamma - \eta \varphi(\overline{I}) - \left(\beta \, h(\overline{I})\right)^{\frac{1}{2}}\right) - F({\overline{I}}),
\end{equation}
and with $F({\overline{I}})$ defined via~\eqref{eq:esse1}. The left-hand side of the algebraic equation~\eqref{eq:exi} is a straight line with positive slope and vertical intercept $0$, whereas the right-hand side is a monotonically decreasing function of $\overline{I}$. Hence, there is a unique $\overline{I}$ such that the algebraic equation~\eqref{eq:exi} is satisfied and for the condition $\overline{I}>0$ to be met it suffices that the right-hand side of equation \eqref{eq:exi} is positive for $\overline{I} \longrightarrow 0^+$. This gives the following condition 
\beq
\label{condposbari}
\gamma -(\eta \beta)^{\frac{1}{2}} - \frac{\nu}{K} \, \left(\xi +\zeta+\zeta \left(\frac{\beta}{\eta}\right)^{\frac{1}{2}}\right)^{-1} > 0.
\eeq
Since the functions $a(x)$ and $b(x)$ are defined according to~\eqref{eq:ab} and assumption~\eqref{ass:noextspec} holds, the basic reproduction number $\mathcal{R}_0$ is given by~\eqref{defR0b} and, therefore,  condition~\eqref{condposbari} is equivalent to the condition $\mathcal{R}_0>1$. 
\end{proof}

\subsection{Long-time asymptotic behaviour}
\label{sec:an3}
A characterisation of the asymptotic behaviour for $t \to \infty$ of the solution to the PIDE-ODE system~\eqref{eq:syst} under assumptions~\eqref{ass:agen} and \eqref{ass:bgen} is provided by the results of Theorem~\ref{th:due}.
\begin{theorem}
\label{th:due}
Under assumptions~\eqref{ass:agen}, \eqref{ass:bgen} and \eqref{ass:noextgen}, the solution to the PIDE-ODE system~\eqref{eq:syst} subject to~\eqref{eq:ICgen}, and to zero Neumann boundary conditions if $X$ is bounded, satisfies the following:
\begin{itemize}
\item[(i)] if 
\beq
\label{cond2g}
\mathcal{R}_0 \leq 1
\eeq
then
\beq
\label{dyngen2a}
s(\cdot,t) \; \longrightarrow \; \overline{s}_F \;\; \text{ in } \;\; L^\infty(X) \; \text{ as } t \to \infty,
\eeq
with $\overline{s}_F(x)$ given by~\eqref{eq:sfgen}, and
\beq
\label{dyngen2b}
I(t) \; \longrightarrow \; 0 \; \text{ as } t \to \infty;
\eeq

\item[(ii)] if 
\beq
\label{cond3g}
\mathcal{R}_0 > 1
\eeq
and either $X$ is bounded or
\begin{equation}
\label{extra_cond}
 b(x) \le B_1 e^{B_2 x} \quad\mbox{for some $B_1, B_2 > 0$}
\end{equation}
then
\beq
\label{dyngen3}
\limsup_{t \to \infty} S(t) > 0 \; \text{ and } \; \limsup_{t \to \infty} I(t) > 0.
\eeq

\end{itemize}
\end{theorem}

\begin{proof} 
{\it Preliminary results used in the proof of (i).} Under assumptions~\eqref{ass:agen}, \eqref{ass:bgen} and~\eqref{ass:noextgen}, we have that $s(\cdot,t) \geq 0$ and $I(t) \geq 0$ for all $t \in [0,\infty)$, and $s(x,t)$ is a sub-solution of the Cauchy problem 
\beq\label{eq:systhat}
\begin{cases}
\displaystyle{\dfrac{\partial \hat{s}}{\partial t} = \left(a(x) - \frac{\hat S}{K}\right) \hat s + \beta \dfrac{\partial^2 \hat{s}}{\partial x^2}}, \quad (x,t) \in X \times (0,\infty),
\\\\
\displaystyle{\hat S(t) := \int_{X} \hat{s}(x,t) \ {\rm d}x},
\\\\
\displaystyle{\hat{s}(0,x) = s^0(x),}
\end{cases}
\quad
\eeq
subject to zero Neumann boundary conditions if $X$ is bounded. It is known (see, for instance, Theorem 3 in~\cite{lorenzi2019asymptotic} for the case where $X$ is bounded) that
\beq
\label{asyshat1}
\hat s(\cdot,t) \; \longrightarrow \; - K \, \lambda \, \psi \;\; \text{ in } \;\; L^\infty(X) \; \text{ as } t \to \infty
\eeq
with $(\lambda,\psi)$ being the principal eigenpair of the elliptic differential operator $\mathcal{L}$ ({\it cf.} Lemma~\ref{lem:zero}). The asymptotic result~\eqref{asyshat1} along with the results established by Theorem~\ref{th:zero} allows us to conclude that
\beq
\label{asyshat2}
\hat{s}(\cdot,t)  \; \longrightarrow \; \overline{s}_F \;\; \text{ in } \;\; L^\infty(X) \; \text{ as } t \to \infty
\eeq
with $\overline{s}_F(x)$ given by~\eqref{eq:sfgen}. The asymptotic result~\eqref{asyshat2} ensures that
\beq
\label{asyRhat}
\dfrac{1}{\nu} \, \int_{X} b(x) \,  \hat{s}(x,t) \, \dd x  \; \longrightarrow \; \mathcal{R}_0 \; \text{ as } t \to \infty
\eeq
with $\mathcal{R}_0$ defined according to~\eqref{defR0gen}. 

\paragraph{Proof of (i).} Assume by contradiction that $I(t)$ is bounded away from zero for $t \to \infty$. If so, $s(x,t)$ is a strict sub-solution of the Cauchy problem~\eqref{eq:systhat} and the asymptotic result~\eqref{asyRhat} along with assumption~\eqref{cond2g} allows one to conclude that there exists $t^* < \infty$ sufficiently large and $\e > 0$ sufficiently small so that
$$
\dfrac{1}{\nu} \, \int_{X} b(x) \,  s(x,t) \, \dd x \leq 1 - \e \quad \text{for all } t \geq t^*.
$$
This along with the ODE~\eqref{eq:syst}$_2$ gives
\beq
\label{neqne}
I' \leq  -\e \, \nu \, I \quad \text{for all } t \geq t^* \quad \Longrightarrow \quad I(t) \; \longrightarrow \;  0  \; \text{ as } t \to \infty,
\eeq
which contradicts the original assumption and allows us to conclude that, under assumption~\eqref{cond2g}, the asymptotic result~\eqref{dyngen2b} holds. Therefore, the PIDE~\eqref{eq:syst}$_1$ can be rewritten as
\beq\label{eq:systrewr}
\displaystyle{\dfrac{\partial s}{\partial t} = \left(a(x) - \frac{S}{K} - \Sigma(x,t)\right) s + \beta \dfrac{\partial^2 s}{\partial x^2}}, 
\eeq
with 
$$
\Sigma(x,t) := \left(b(x) + \dfrac{1}{K}\right) I(t).
$$
Due to the structure of the ODE~\eqref{eq:syst}$_2$, the asymptotic result~\eqref{dyngen2b} implies that $\Sigma(x,t) \longrightarrow 0$ exponentially fast in $L^\infty(X)$ as $t \to \infty$. As a result, starting from~\eqref{eq:systrewr} and using arguments analogous to those used to obtain the asymptotic result~\eqref{asyshat2} one can prove that the asymptotic result~\eqref{dyngen2a} holds.

\paragraph{Preliminary results used in the proof of (ii).} Integrating both sides of the PIDE~\eqref{eq:syst}$_1$ over $X$ and adding the resulting ODE for $S(t)$ and the ODE~\eqref{eq:syst}$_2$ for $I(t)$ we obtain the following ODE for $N(t) := S(t)+I(t)$
$$
N' = \int_{X} a(x) s(x,t) \ {\rm d}x - \nu I  - \frac{N}{K} S.
$$
Estimating from above using the nonnegativity of $S$ and $I$, the positivity of $\nu$ and the fact that $\displaystyle{\max_{x \in \mathbb{R}} a(x) = a(0)>0}$ (cf. assumptions~\eqref{ass:agen}) we obtain the following differential inequality
$$
N' \leq \left(a(0) - \dfrac{N}{K}\right) S.
$$
Since $N(0)<\infty$ (cf. initial conditions~\eqref{eq:ICgen}), $a(0)<\infty$ and $0 \le S\le N$ is nonnegative, the latter differential inequality implies that
\beq
\label{UBN}
N(t) \le \max\{N(0), K a(0)\} <\infty \quad \text{for all } t \in [0,\infty).
\eeq

\paragraph{Proof of (ii).} Assume by contradiction that $I(t) \longrightarrow 0$ as $t \to \infty$. If this occurs, then the PIDE~\eqref{eq:syst}$_1$ can be rewritten in the form of~\eqref{eq:systrewr} and the same arguments used in the proof of (i) yield 
$$
s(\cdot,t) \; \xrightarrow[t\rightarrow\infty] \; \overline{s}_F \;\; \text{ in } \;\; L^\infty(X).
$$
The latter asymptotic result along with assumption~\eqref{cond3g} allows one to conclude that there exists $t^* < \infty$ sufficiently large and $\e > 0$ sufficiently small so that
$$
\dfrac{1}{\nu} \, \int_{X} b(x) \,  s(x,t) \, \dd x > 1+ \e \quad \text{for all } t \geq t^*.
$$
This along with the ODE~\eqref{eq:syst}$_2$ gives 
$$
I' \geq  I \, \e \, \nu \quad \text{for all } t \geq t^* \quad \Longrightarrow \quad I(t) \; \longrightarrow \; \infty  \; \text{ as } t \to \infty,
$$
which contradicts the assumption that $I(t) \longrightarrow 0$ as $t \to \infty$, as well as the upper bound~\eqref{UBN}. Since $I(t) \ge 0$, we conclude $\displaystyle{\limsup_{t \to \infty} I(t) > 0}$. The structure of the ODE~\eqref{eq:syst}$_2$ immediately implies that 
\begin{equation}
\label{lim_bx}
 \limsup_{t \to \infty}  \int_{X} b(x) \,  s(x,t) \, \dd x \ge  \nu  .
\end{equation}

If $X \equiv \mathbb{R}$, since 
$$ 
s(x,t) \le \hat s(x,t) \quad \text{with} \quad \hat s(x,t) \; \xrightarrow[t\rightarrow\infty] \;\; \overline{s}_F(x) = \overline{S}_F \, \psi(x) 
$$
and assumptions~\eqref{ass:agen} imply \cite[Corollary 2.3.2.1]{berezinshubin} that for any $k > 0$ there exists $A > 0$ such that 
$$ 
\psi(x) \le e^{- k |x|}\quad\mbox{for }|x| \ge A, 
$$
assumption~\eqref{extra_cond} allows us to conclude that for each $\e > 0$ there exist $A>0$ and $t^*<\infty$ such that
$$ \int_{|x| > A} b(x) \,  s(x,t) \, \dd x \le \e \quad\forall\ t \ge t^*. $$
Hence,
\begin{equation}
\label{conc}
\limsup_{t \to \infty} S(t) \ge \limsup_{t \to \infty}  \int\limits_{|x| \le  A}  \,  s(x,t) \, \dd x  \ge \limsup_{t \to \infty} \dfrac{\displaystyle{\int\limits_{|x| \le A} b(x) \,  s(x,t) \, \dd x}}{\max\limits_{|x| \le A} b(x) } > 0.
\end{equation}

If $X$ is bounded then
$$ 
S(t) \ge \dfrac{1}{\displaystyle{\max_{x \in X} b(x)}} \displaystyle{\int_{X} b(x) \,  s(x,t) \, \dd x}
$$
and the conclusion follows immediately from \eqref{lim_bx}.
\end{proof}
\begin{remark}
Notice that the asymptotic result~\eqref{dyngen3} is what is called weak persistence in~\cite{Smith2011c}. It seems likely that more careful estimates could lead to uniform weak persistence so that the use of appropriate theorems from Chapter 4 of~\cite{Smith2011c} could lead to uniform strong persistence. This is certainly possible in the case when definitions~\eqref{eq:ab} are considered, because in this case the PIDE~\eqref{eq:syst}$_1$ can be reduced to a finite-dimensional system, but it is not pursued here.
\end{remark}

\section{Numerical results and biological interpretation}
\label{Numerics}
In this section, we report on the results of numerical simulations that complement the analytical results established in Section~\ref{Analytics}. In Section~\ref{Numerics0}, model calibration, set-up of numerical simulations and numerical methods are described. In Section~\ref{Numerics1}, the results of base-case numerical simulations carried out assuming either $\mathcal{R}_0 \leq 1$ or $\mathcal{R}_0 > 1$ are presented. In Section~\ref{Numerics2}, these results are integrated with the results of additional numerical simulations that summarise the outcomes of a systematic sensitivity analysis with respect to the model parameters, and the biological implications of our theoretical findings are discussed.

\subsection{Model calibration, set-up of numerical simulations and numerical methods}
\label{Numerics0}
\paragraph{Model calibration.} We define the functions $a(x)$ and $b(x)$ via~\eqref{eq:ab}. The parameter values used to carry out base-case numerical simulations are listed in Table~\ref{table1}. These parameters are estimated using the results of artificial selection experiments in a host-parasite system presented by Webster \& Woolhouse~\cite{WeWo}, who demonstrated the existence of a fitness trade-off between resistance to parasitic {\it Schistosoma mansoni} and fertility in {\it Biomphalaria glabrata}. In summary, the values of the parameters $\gamma$ and $\eta$ are chosen such that the intrinsic net per capita growth rates of individuals in the phenotypic states $x=0$ and $x=1$ (i.e. the values of $a(0) = \gamma$ and $a(1) = \gamma - \eta$) match with the average experimental net proliferation rates of host individuals with high fertility and high resistance to infection, respectively. Notice that the values of $\gamma$ and $\eta$ so obtained are such that, coherently with assumptions~\eqref{ass:agen}, we have $\gamma > \eta$.
Given the values of $\gamma$ and $\eta$, we identified the values of the parameters $\xi$, $\zeta$ and $K$ through exploratory numerical simulations carried out assuming $\beta=0$ (i.e. neglecting phenotypic changes), letting the value of the parameter $\nu$ match the experimental value of the death rate of infected individuals, and choosing $\xi$, $\zeta$ and $K$ in such a way as to minimise the mean square error between the average experimental value of the number of infected individuals at different time instants and the corresponding values of $I(t)$. Finally, the value of the parameter $\beta$ is chosen such that condition~\eqref{ass:noextspec} is met and the value of $\nu$ is chosen on a case-by-case basis so that $\mathcal{R}_0$ given by~\eqref{defR0b} satisfies either~\eqref{cond2g} or \eqref{cond3g}.

\begin{remark}
Since the functions $a(x)$ and $b(x)$ are defined via~\eqref{eq:ab}, throughout the rest of the paper the basic reproduction number $\mathcal{R}_0$ is computed via~\eqref{defR0b}.
\end{remark}

\renewcommand{\arraystretch}{1.2}
\begin{table}[h!]
	\begin{center}
	\begin{tabular}{l|l|l}
		Parameter                        & Description                                                                                   & Values                                          \\ \hline
		\multicolumn{1}{|l|}{$\gamma$}   & Maximum intrinsic net per capita growth rate                                                                                             & \multicolumn{1}{l|}{0.2645}                     \\ \hline
		\multicolumn{1}{|l|}{$\eta$}     & Intrinsic selection gradient                                                                                      & \multicolumn{1}{l|}{0.1945}                    \\ \hline
		\multicolumn{1}{|l|}{$\nu$}      & \begin{tabular}[c]{@{}l@{}}Rate of death caused by infection\end{tabular}                  & \multicolumn{1}{l|}{$\left\{0.2, 2 \right\}$}                  \\ \hline   
		\multicolumn{1}{|l|}{$\xi$} & Minimum infection rate                                                                                      & \multicolumn{1}{l|}{1.5219 $\times 10^{-5}$} \\ \hline
		\multicolumn{1}{|l|}{$\zeta$}    & \begin{tabular}[c]{@{}l@{}}Selection gradient related to infection\end{tabular} & \multicolumn{1}{l|}{6 $\times 10^{-3}$}                   \\ \hline
		\multicolumn{1}{|l|}{$K$}        & Rescaled carrying capacity                                                                                      & \multicolumn{1}{l|}{$10^3$}                    \\ \hline
		\multicolumn{1}{|l|}{$\beta$}    & \begin{tabular}[c]{@{}l@{}}Rate of spontaneous phenotypic variation\end{tabular}                          & \multicolumn{1}{l|}{$0.01$}                     \\ \hline
	\end{tabular}
\end{center}
	\caption{Values of the parameters used to carry out base-case numerical simulations. The parameter $K$ is in units of $number \ of \ individuals \times week$, while all the other parameters are in units of $week^{-1}$.}
\label{table1}
\end{table}

\paragraph{Set-up of numerical simulations and numerical methods.} We select a uniform discretisation consisting of $1200$ points on the interval $(-L,L)$ with $L=10$ as the computational domain of the independent variable $x$ and impose zero Neumann boundary conditions in $x=-L$ and $x=L$. Moreover, we assume $t \in (0,T]$, with $T>0$ being the final time of simulations, and we discretise the interval $(0,T]$ with the uniform step $\Delta t= 10^{-4}$. The method for constructing numerical solutions to the PIDE~\eqref{eq:syst}$_1$ is based on an explicit finite difference scheme in which a three-point stencil is used to approximate the diffusion term and an explicit finite difference scheme is used for the reaction term~\cite{leveque2007finite}. Numerical solutions to the ODE~\eqref{eq:syst}$_2$ are constructed using the explicit Euler method. All numerical computations are performed in {\sc Matlab}. We consider an initial condition that satisfies~\eqref{eq:ICgen}, that is,
$$
s(x,0) =  10 \ \frac{c^0}{\left(2 \pi\right)^{\frac{1}{2}}} \, \exp\left[-\frac{1}{2} \left(x - 0.5 \right)^2\right] \quad \text{and} \quad I(0) = 10
$$
with $c^0 \in \mathbb{R}^+_{*}$ being a normalisation constant such that 
$$
 \frac{c^0}{\left(2 \pi\right)^{\frac{1}{2}}} \int_{-L}^L \exp\left[-\frac{1}{2} \left(x - 0.5 \right)^2\right] \ {\rm d}x = 1.
$$

\subsection{Base-case numerical simulations}
\label{Numerics1}

\begin{figure}[h!]
	\begin{center}
		\includegraphics[width=1\linewidth]{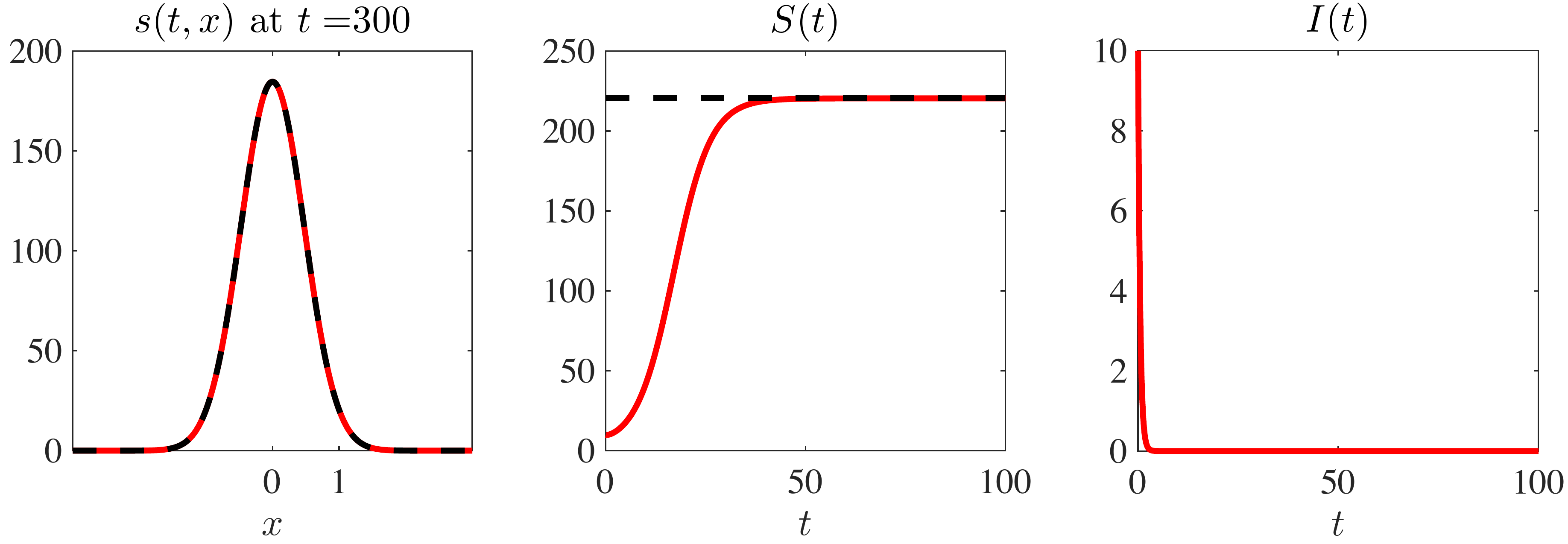}
		\caption{{\bf Base-case numerical simulations for $\mathcal{R}_0 \leq 1$.} Sample dynamics of the size of the susceptible compartment $S(t)$ (central panel, solid line) and the number of infected individuals $I(t)$ (right panel) in the case where condition~\eqref{cond2g} is met  (i.e. when $\mathcal{R}_0 \leq 1$). The dashed line in the central panel highlights the equilibrium value $\overline{S}_F$ given by~\eqref{eq:st1barS}. The population density function $s(x,t)$ at the final time $t=300$ is displayed in the left panel (solid line), where the dashed line highlights the equilibrium population density function $\overline{s}_F(x)$ given by~\eqref{eq:st1}.
		The values of the model parameters are those reported in Table~\ref{table1} with $\nu=2$.}
		\label{fig2}
	\end{center}
\end{figure}

\paragraph{Base-case numerical simulations for $\mathcal{R}_0 \leq 1$.} The numerical solutions displayed in Figure~\ref{fig2} demonstrate that, in agreement with the asymptotic results of claim (i) in Theorem~\ref{th:due}, when condition~\eqref{cond2g} is met (i.e. if $\mathcal{R}_0 \leq 1$) $I(t)$ decays to zero while $s(x,t)$ converges to an equilibrium population density function. Coherently with Theorem~\ref{th:zero} and Proposition~\ref{prop:zero}, the equilibrium population density function is the function $\overline{s}_F(x)$ given by~\eqref{eq:st1} and, therefore, $S(t)$ converges to the equilibrium value $\overline{S}_F$ given by~\eqref{eq:st1barS}.

\begin{figure}[h!]
	\begin{center}
		\includegraphics[width=1\linewidth]{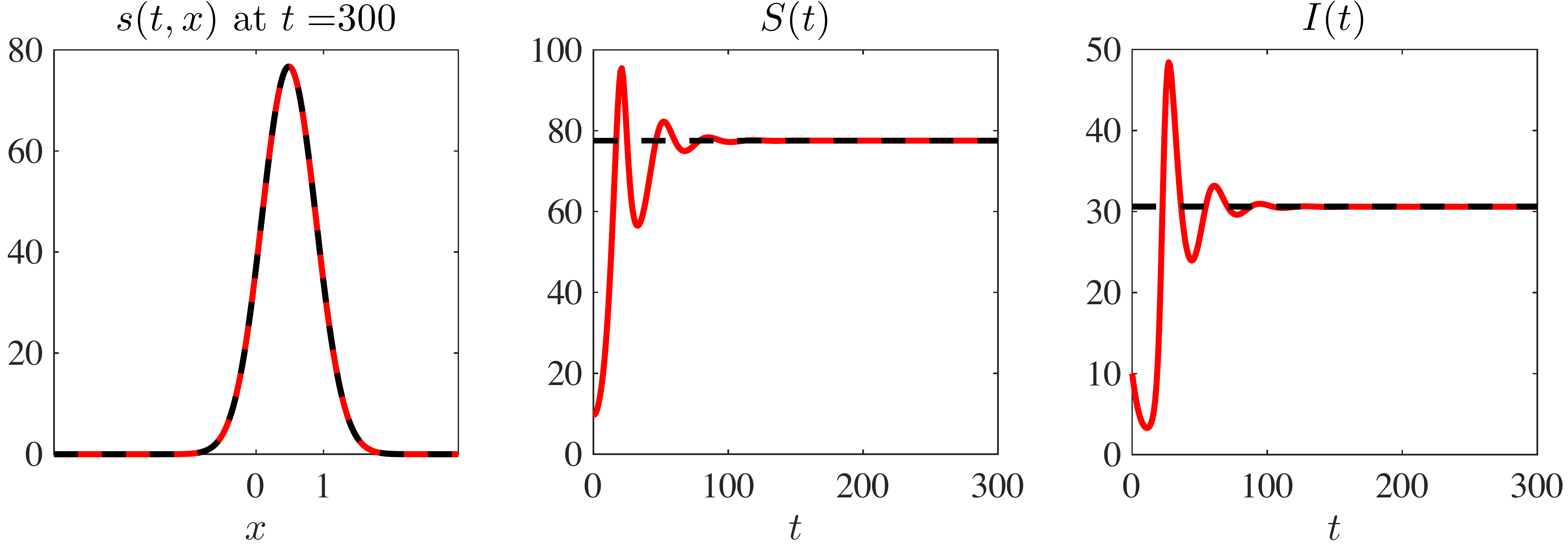}
		\caption{{\bf Base-case numerical simulations for $\mathcal{R}_0 > 1$.} Sample dynamics of the size of the susceptible compartment $S(t)$ (central panel, solid line) and the number of infected individuals $I(t)$ (right panel, solid line) in the case where condition~\eqref{cond3g} is met (i.e. when $\mathcal{R}_0 > 1$). The dashed line in the right panel highlights the value of $\overline{I}_E$ obtained by  solving numerically the algebraic equation~\eqref{eq:equilI}, while the dashed line in the central panel highlights the value of $\overline{S}_E$ given by~\eqref{eq:equilnbarS}. The population density function $s(x,t)$ at the final time $t=300$ is displayed in the left panel (solid line), where the dashed line highlights the equilibrium population density function $\overline{s}_E(x)$ given by~\eqref{eq:equiln}. The values of the model parameters are those reported in Table~\ref{table1} with $\nu=0.2$.}
		\label{fig3}
	\end{center}
\end{figure}

\paragraph{Base-case numerical simulations for $\mathcal{R}_0 > 1$.} The numerical solutions displayed in Figure~\ref{fig3} demonstrate that, in agreement with the asymptotic results of claim (ii) in Theorem~\ref{th:due}, when condition~\eqref{cond3g} is met (i.e. if $\mathcal{R}_0 \leq 1$) both $I(t)$ and $S(t)$ remain bounded away from zero. It can also be seen that $I(t)$ converges to a stationary value, which is the unique equilibrium $\overline{I}_E$, the existence of which is shown in Theorem~\ref{th:uno}, which is computed by solving numerically the algebraic equation~\eqref{eq:equilI}. Moreover, $s(x,t)$ converges to the equilibrium population density function $\overline{s}_E(x)$ given by~\eqref{eq:equiln} and, therefore, $S(t)$ converges to the equilibrium value $\overline{S}_E$ given by~\eqref{eq:equilnbarS}.

\subsection{Sensitivity analysis with respect to the model parameters and main biological implications}
\label{Numerics2}
\paragraph{Conditions underpinning spread of infection.} The results of claim (i) in Theorem~\ref{th:due}, along with the numerical solutions displayed in Figure~\ref{fig2}, demonstrate that if $\mathcal{R}_0 \leq 1$ then an infection cannot spread and a disease-free equilibrium is ultimately attained. On the other hand, the results of claim (ii) in Theorem~\ref{th:due}, along with the numerical solutions displayed in Figure~\ref{fig3}, show that if $\mathcal{R}_0 > 1$ then the infection will spread. 

\begin{figure}[h!]
	\begin{center}
		\includegraphics[width=1\linewidth]{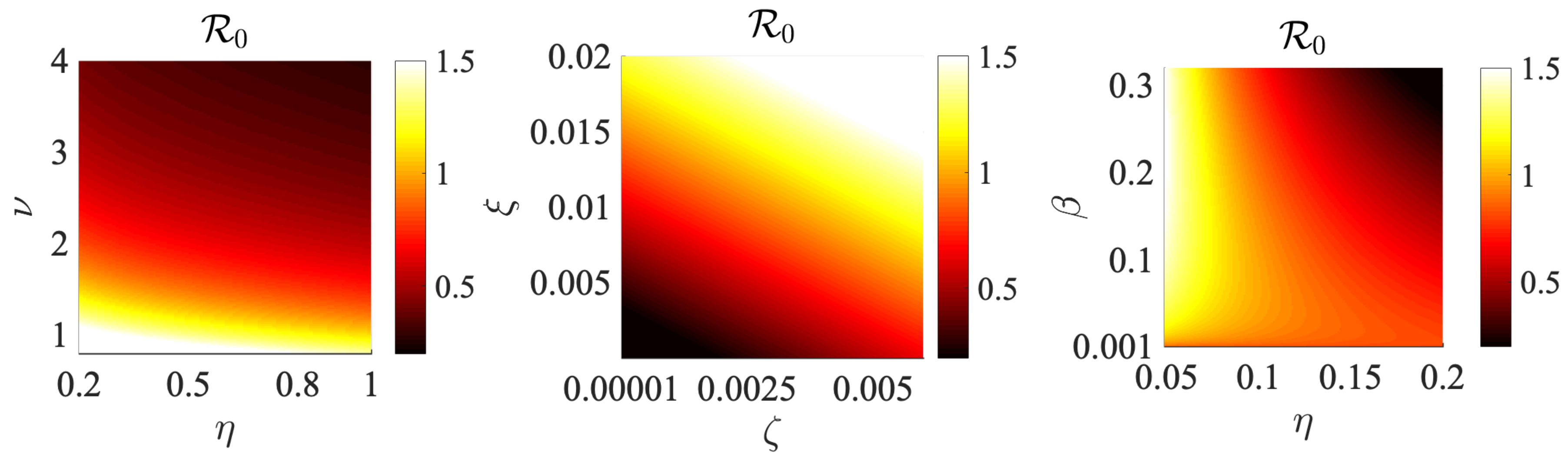}
		\caption{{\bf Conditions underpinning spread of infection.} Plots of the basic reproduction number $\mathcal{R}_0$ as a function of $\eta$ and $\nu$ (left panel), as a function of $\zeta$ and $\xi$ (central panel), and as a function of $\eta$ and $\beta$ (right panel). The values of the other model parameters are those reported in Table~\ref{table1} with $\nu=2$ for the plots in the central and right panels, while $\beta=0.1$ for the plot in the central panel.}
		\label{fig4}
	\end{center}
\end{figure}

From~\eqref{defR0b} one sees that the basic reproduction number $\mathcal{R}_0$ is a decreasing function of the rate of death caused by infection $\nu$ and of the intrinsic selection gradient $\eta$, and an increasing function of the minimum infection rate $\xi$ and of the selection gradient related to infection $\zeta$. Moreover, $\mathcal{R}_0$ is an increasing function of $\beta$ for $\eta$ sufficiently low and a decreasing function of $\beta$ for $\eta$ sufficiently high. This is illustrated by the heat maps in Figure~\ref{fig4}.

From the disease-free equilibrium given in Proposition~\ref{prop:zero}, one sees that the asymptotic value of the size of the susceptible compartment $S$ at the disease-free equilibrium and the basic reproduction number $\mathcal{R}_0$ are both decreasing in $\eta$. Moreover, from the endemic equilibrium given in Proposition~\ref{prop:uno}, we observe that the asymptotic value of the mean phenotypic state $\mu$, bounded from below by $0$, is decreasing in $\eta$ as well, meaning that the higher is $\eta$ the smaller the susceptible population will be, and it will be concentrated on a phenotypic range which makes it more vulnerable to the infection.

These results provide a mathematical formalisation of the idea that infections exerting stronger selective pressures on susceptible individuals, and being characterised by lower mortality rates and higher infection rates, are more likely to spread. Furthermore, when susceptible individuals are exposed to weaker intrinsic selective pressures, frequent heritable, spontaneous phenotypic changes may promote the spread of an infection. On the other hand, higher rates of heritable, spontaneous phenotypic changes may correlate with a lower risk of spreading infection in the case where susceptible individuals are subject to stronger intrinsic selective pressures.

\begin{figure}[h!]
	\begin{center}
		\includegraphics[width=1\linewidth]{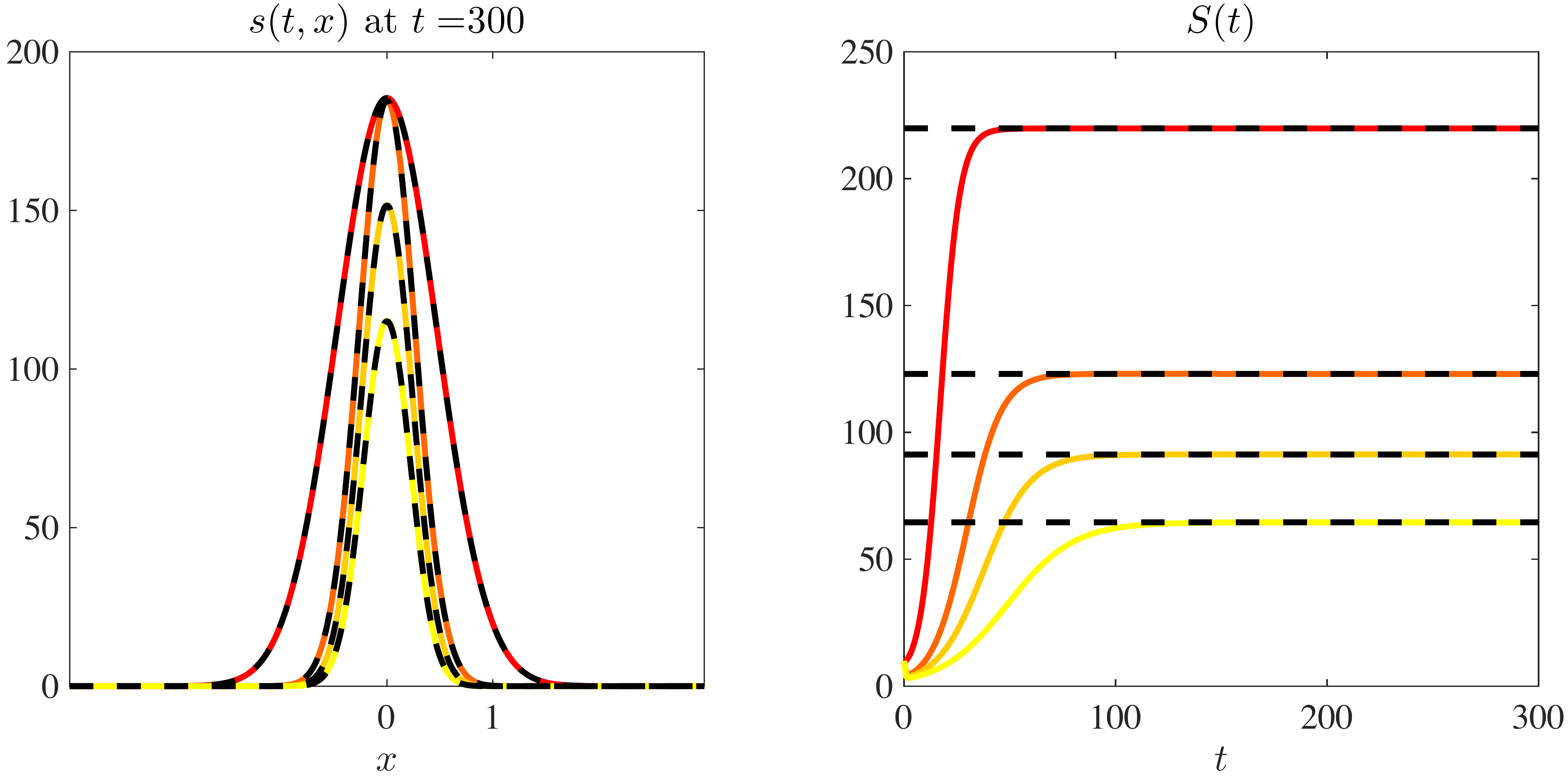}
		\caption{{\bf Dependence of the disease-free equilibrium on evolutionary parameters.} Sample dynamics of the size of the susceptible compartment $S(t)$ (right panel, solid lines) in the case where condition~\eqref{cond2g} is met (i.e. when $\mathcal{R}_0 \leq 1$) and for increasing values of $\eta$, that is, $\eta=0.2$ (red line), $\eta=2$ (orange line), $\eta=3$ (light orange line) and $\eta=4$ (yellow line). The dashed lines in the right panel highlight the value of $\overline{S}_F$ given by~\eqref{eq:st1barS}. The corresponding population density functions $s(x,t)$ at the final time $t=300$ are displayed in the left panel (solid lines), where the dashed lines highlight the equilibrium population density function $\overline{s}_F(x)$ given by~\eqref{eq:st1}. The values of the other model parameters are those reported in Table~\ref{table1} with $\nu=2$.}
		\label{fig5a}
	\end{center}
\end{figure}

\begin{figure}[h!]
	\begin{center}
		\includegraphics[width=1\linewidth]{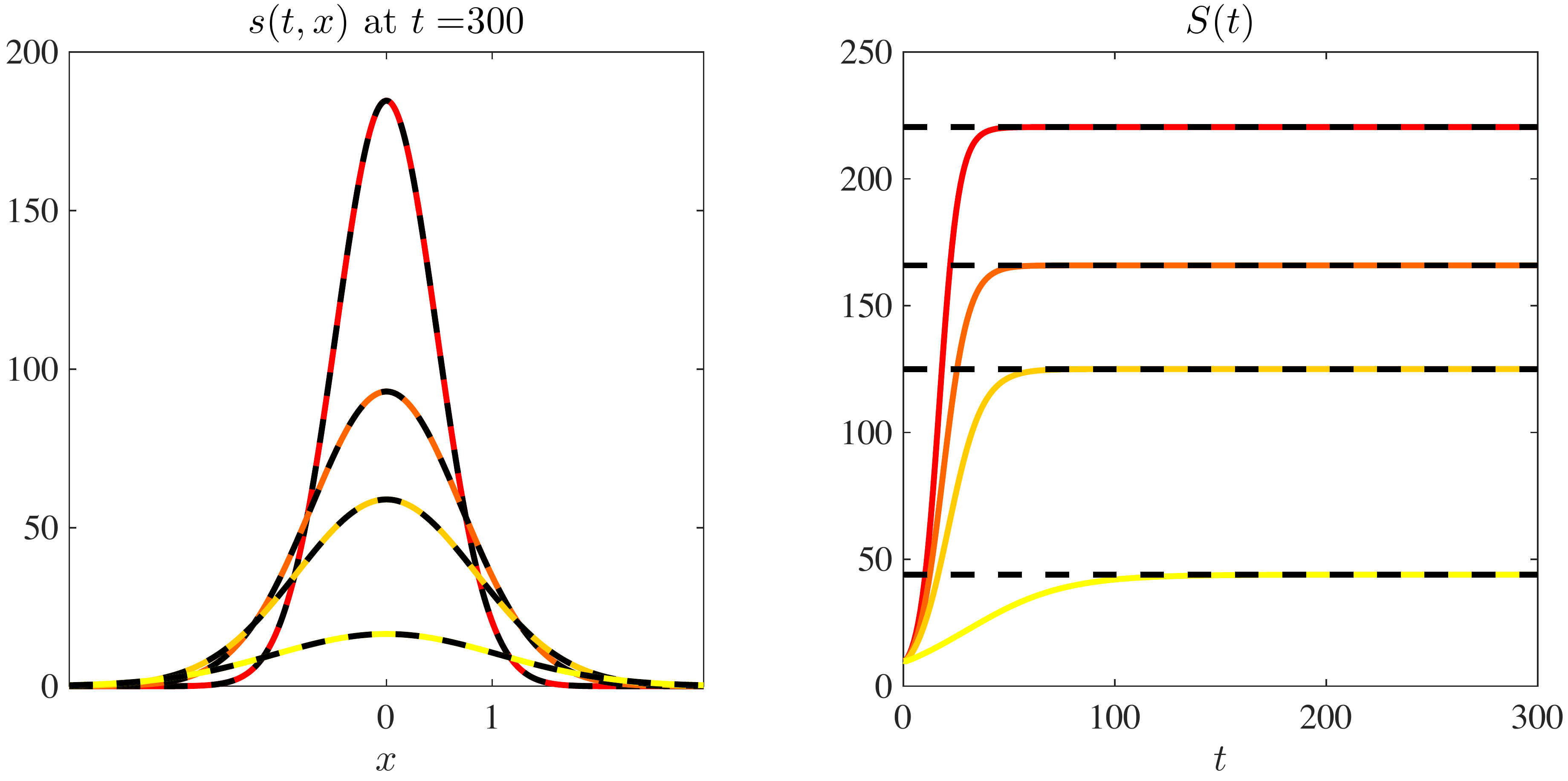}
		\caption{{\bf Dependence of the disease-free equilibrium on evolutionary parameters.} Sample dynamics of the size of the susceptible compartment $S(t)$ (right panel, solid lines) in the case where condition~\eqref{cond2g} is met (i.e. when $\mathcal{R}_0 \leq 1$) and for increasing values of $\beta$, that is, $\beta=0.01$ (red line), $\beta=0.05$ (orange line), $\beta=0.1$ (light orange line) and $\beta=0.25$ (yellow line). The dashed lines in the right panel highlight the value of $\overline{S}_F$ given by~\eqref{eq:st1barS}. The corresponding population density functions $s(x,t)$ at the final time $t=300$ are displayed in the left panel (solid lines), where the dashed lines highlight the equilibrium population density function $\overline{s}_F(x)$ given by~\eqref{eq:st1}. The values of the other model parameters are those reported in Table~\ref{table1} with $\nu=2$.}
		\label{fig5b}
	\end{center}
\end{figure}

\paragraph{Dependence of the disease-free equilibrium on evolutionary parameters.} Coherently with Theorem~\ref{th:zero} and Proposition~\ref{prop:zero}, the numerical solutions presented in Figure~\ref{fig2} indicate that, if a disease-free equilibrium is attained, the size of the susceptible compartment $S$ converges to the equilibrium value $\overline{S}_F$ given by~\eqref{eq:st1barS}, which increases with the maximum intrinsic net per capita growth rate $\gamma$ and the rescaled carrying capacity $K$. Moreover, as shown also by the numerical solutions displayed in Figure~\ref{fig5a} and Figure~\ref{fig5b}, the equilibrium value $\overline{S}_F$ decreases with the intrinsic selection gradient $\eta$ and the rate of heritable, spontaneous phenotypic variation $\beta$. 

The phenotypic distribution of susceptible individuals at disease-free equilibrium is of the Gaussian type~\eqref{eq:st1}. The mean value of the equilibrium phenotypic distribution corresponds to the fittest phenotypic state $\varphi(0)$ defined via~\eqref{ghori}, which coincides with the phenotypic state characterised by the highest proliferative potential, i.e.
$$
\mu(t) \longrightarrow \varphi(0) = 0 \quad \text{as } t \to \infty.
$$
The variance of the equilibrium phenotypic distribution is governed by the ratio between $\beta$ and the effective selection gradient $h(0)$ defined via~\eqref{ghori}, which coincides with the intrinsic selection gradient $\eta$, i.e.
$$
\sigma^2(t) \longrightarrow \left(\dfrac{\beta}{h(0)}\right)^{\frac{1}{2}} = \left(\dfrac{\beta}{\eta}\right)^{\frac{1}{2}} \quad \text{as } t \to \infty.
$$
This is confirmed by the numerical solutions displayed in Figure~\ref{fig5a} and Figure~\ref{fig5b}.

These results demonstrate that, when a disease-free equilibrium is attained, the susceptible population will be mainly composed of individuals with a high proliferative potential and the degree of phenotypic heterogeneity will be an increasing function of the rate of heritable, spontaneous phenotypic variation and a decreasing function of the intrinsic selection gradient. 

\begin{figure}[h!]
	\begin{center}
		\includegraphics[width=1\linewidth]{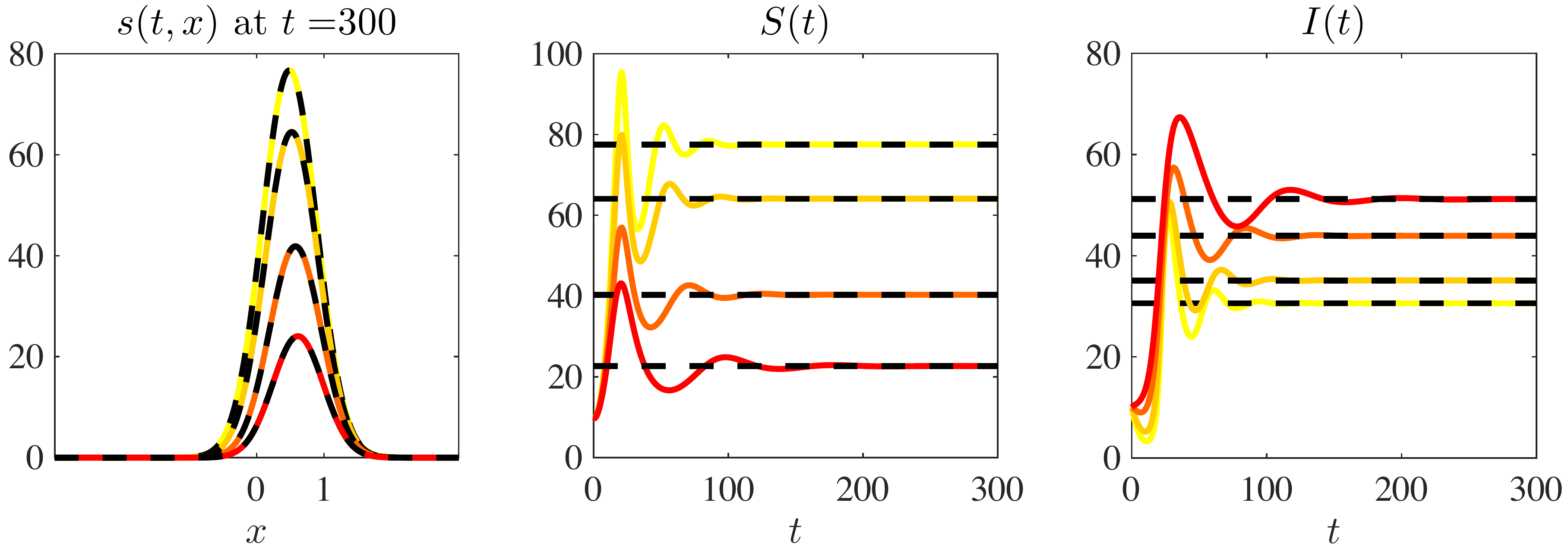}
		\caption{{\bf Dependence of the endemic equilibrium on evolutionary parameters.} Sample dynamics of the size of the susceptible compartment $S(t)$ (central panel, solid lines) and the number of infected individuals $I(t)$ (right panel, solid lines) in the case where condition~\eqref{cond3g} is met (i.e. when $\mathcal{R}_0>1$) and for decreasing values of $\nu$, that is, $\nu=0.2$ (yellow lines), $\nu=0.15$ (light orange lines), $\nu=0.08$ (orange lines) and $\nu=0.04$ (red lines). The dashed lines in the right panel highlight the values of $\overline{I}_E$ obtained by solving numerically the algebraic equation~\eqref{eq:equilI}, while the dashed lines in the central panel highlight the values of $\overline{S}_E$ given by~\eqref{eq:equilnbarS}. The corresponding population density functions $s(x,t)$ at the final time $t=300$ are displayed in the left panel (solid lines), where the dashed lines highlight the equilibrium population density function $\overline{s}_E(x)$ given by~\eqref{eq:equiln}. The values of the other model parameters are those reported in Table~\ref{table1}.}
		\label{fig6}
	\end{center}
\end{figure}

\begin{figure}[h!]
	\begin{center}
		\includegraphics[width=1\linewidth]{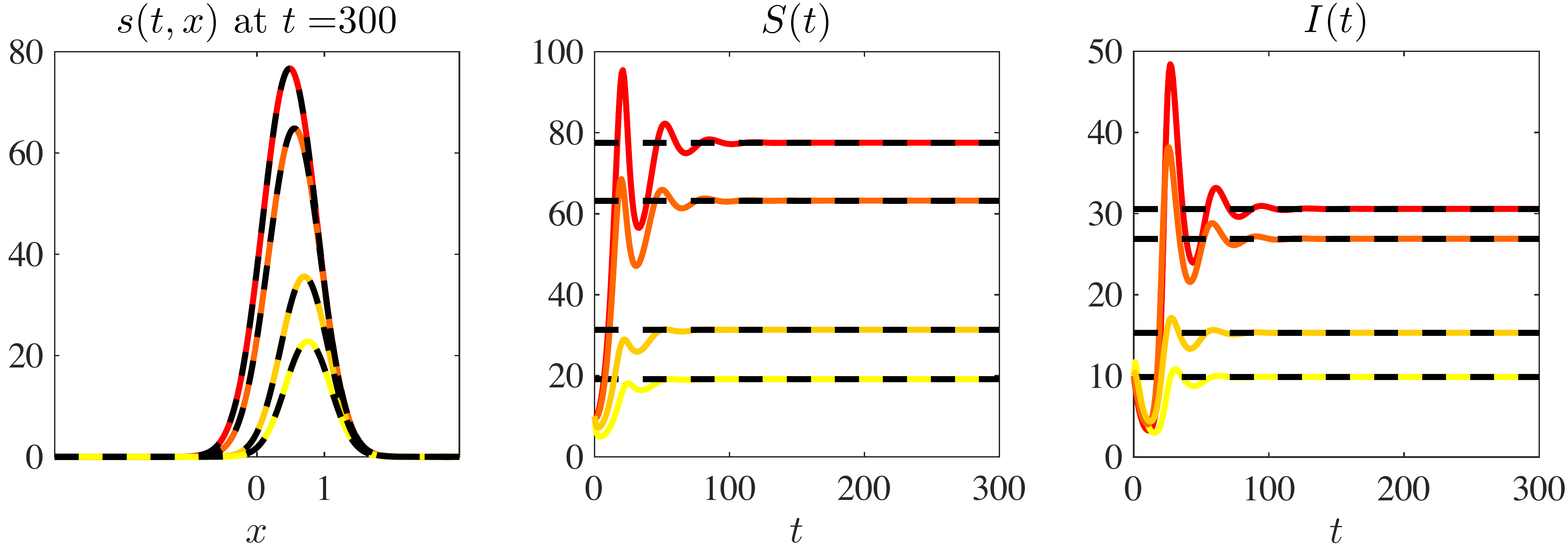}
		\caption{{\bf Dependence of the endemic equilibrium on evolutionary parameters.} Sample dynamics of the size of the susceptible compartment $S(t)$ (central panel, solid lines) and the number of infected individuals $I(t)$ (right panel, solid lines) in the case where condition~\eqref{cond3g} is met (i.e. when $\mathcal{R}_0>1$) and for increasing values of $\zeta$, that is, $\zeta=0.006$ (red lines), $\zeta=0.009$ (orange lines), $\zeta=0.03$ (light orange lines) and $\zeta=0.06$ (yellow lines). The dashed lines in the right panel highlight the values of $\overline{I}_E$ obtained by solving numerically the algebraic equation~\eqref{eq:equilI}, while the dashed lines in the central panel highlight the values of $\overline{S}_E$ given by~\eqref{eq:equilnbarS}. The corresponding population density functions $s(x,t)$ at the final time $t=300$ are displayed in the left panel (solid lines), where the dashed lines highlight the equilibrium population density function $\overline{s}_E(x)$ given by~\eqref{eq:equiln}. The values of the other model parameters are those reported in Table~\ref{table1} with $\nu=0.2$.}
		\label{fig7}
	\end{center}
\end{figure}

\paragraph{Dependence of the endemic equilibrium on evolutionary parameters.} Coherently with Theorem~\ref{th:uno} and Proposition~\ref{prop:uno}, the numerical solutions presented in Figure~\ref{fig3} indicate that, if an endemic equilibrium is attained, the number of infected individuals $I$ converges to the unique positive solution $\overline{I}_E$ of the algebraic equation~\eqref{eq:equilI}, while the size of the susceptible compartment $S$ converges to the equilibrium value $\overline{S}_E$ given by~\eqref{eq:equilnbarS}. Since $\overline{I}_E$ is a monotonically decreasing function of $\nu$ and $\overline{S}_E$ is a monotonically decreasing function of $\overline{I}_E$, we have that $\overline{S}_E$ is a monotonically increasing function of $\nu$. This is confirmed by the numerical solutions displayed in Figure~\ref{fig6}. Moreover, the numerical results summarised by Figure~\ref{fig7} show that, for the choice of parameter values considered here, both $\overline{I}_E$ and $\overline{S}_E$ decrease with $\zeta$.

The phenotypic distribution of susceptible individuals at endemic equilibrium is of the Gaussian type~\eqref{eq:equiln}. The mean value of the equilibrium phenotypic distribution corresponds to the fittest phenotypic state $\varphi(\overline{I}_E)$ defined via~\eqref{ghori}, i.e.
$$
\mu(t) \longrightarrow \varphi(\overline{I}_E) = \dfrac{\zeta \overline{I}_E}{\eta + \zeta \overline{I}_E} \quad \text{as } t \to \infty.
$$
The variance of the equilibrium phenotypic distribution is given by the ratio between $\beta$ and the effective selection gradient $h(\overline{I}_E)$ defined via~\eqref{ghori}, i.e.
$$
\sigma^2(t) \longrightarrow \left(\dfrac{\beta}{h(\overline{I}_E)}\right)^{\frac{1}{2}} = \left(\frac{\beta}{\eta + \zeta \overline{I}_E}\right)^{\frac{1}{2}} \quad \text{as } t \to \infty.
$$ 
Hence, larger values of $\zeta$ and smaller values of $\nu$ -- since $\overline{I}_E$ is a monotonically decreasing function of $\nu$, as mentioned earlier -- will lead the peak of the equilibrium phenotype distribution of susceptible individuals to move from the phenotypic state characterised by the highest proliferative potential (i.e. $x=0$) toward phenotypic states closer to the phenotypic state corresponding to the highest level of resistance to infection (i.e. $x=1$). Moreover, higher $\zeta$ and smaller $\nu$ will correlate with a narrower equilibrium phenotype distribution of susceptible individuals (i.e. lower degrees of phenotypic heterogeneity in the susceptible compartment). This is confirmed by the numerical solutions presented in Figure~\ref{fig6} and Figure~\ref{fig7}.

These results support the idea that, if an endemic equilibrium is established, infections characterised by lower rates of death and exerting stronger selective pressures on susceptible individuals will lead to a smaller and less phenotypically diverse susceptible compartment, which will be mainly composed of individuals with lower proliferative potential and higher level of resistance to infection.

\section{Conclusions and research perspectives}
\label{Conclusions} 
The results of our mathematical study disentangle the impact of different evolutionary parameters on the spread of infectious diseases and the consequent phenotypic adaption of susceptible individuals. In particular, the results obtained provide a theoretical basis for the observation that infectious diseases exerting stronger selective pressures on susceptible individuals and being characterised by higher infection rates are more likely to spread. Furthermore, these results indicate that heritable, spontaneous phenotypic changes in proliferative potential and resistance to infection can either promote or prevent the spread of infectious diseases depending on the strength of selection acting on susceptible individuals prior to infection. Finally, we have demonstrated that, when an endemic equilibrium is established, higher levels of resistance to infection and lower degrees of phenotypic heterogeneity among susceptible individuals are to be expected in the presence of infections which are characterised by lower rates of death and exert stronger selective pressures on susceptible individuals. 
Note that such results have been obtained without making any smallness assumption on the rate of phenotypic changes. This strengthens the robustness of the biological inferences drawn from the mathematical results obtained.
 
As a complement to the analytical part of our work, we chose some more specific biologically relevant definitions of the model functions and, by using a suitable ansatz and carrying out direct computations, we provided a fully-explicit characterisation of the disease-free equilibrium and endemic equilibrium of the model. We believe that the results obtained showcase the application potential of the analyses undertaken in the present study.

We conclude with an outlook on possible research perspectives. From a mathematical point of view, it would be interesting to complement the asymptotic results established by Theorem~\ref{th:due} by proving convergence to the endemic equilibrium when condition $\mathcal{R}_0 > 1$ is satisfied. This will require to find a suitable Lyapunov function for the PIDE-ODE system~\eqref{eq:syst}. Moreover, it would be interesting to explore cases where infection transmission is frequency-dependent, instead of being density-dependent, and cases where a phenotypic structure is introduced in the infective compartment. It would also be interesting to consider the more general case of SIR models, which will make it necessary to introduce a phenotypic structure both in the infected and recovered compartments, in order to take into account the effect of the reproduction of recovered individuals on the evolutionary dynamics of the system. All of these extensions would require further development of the methods of proof presented here in order to carry out similar analyses of evolutionary dynamics. 

Since the analytical results we have obtained can accommodate parameter values for a wide range of SI systems and a variety of infectious diseases, from an application point of view it would be interesting to apply these results to specific datasets in order to test the validity of the explicit condition for the spread of infection that we have derived. Moreover, building upon previous work on the derivation of deterministic continuum models for the evolution of populations structured by phenotypic traits from stochastic individual-based models~\cite{ardavseva2020comparative,champagnat2002canonical,champagnat2006unifying,chisholm2016evolutionary,stace2020discrete}, another track to follow to further enrich the present study would be to develop a stochastic individual-based model corresponding to the deterministic continuum model presented here. This would make it possible to explore the impact of stochastic fluctuations in phenotypic properties of single individuals on the spread of infections. 

Finally, given the fact that spatial interactions are central to the spread of several infectious diseases, an additional natural way of extending our study would be to introduce spatial structure in order to consider scenarios in which individuals move across space and/or migrate between different regions that occupy the nodes of a network.

\section*{Acknowledgements}
T.L. gratefully acknowledges the hospitality provided by the Department of Mathematics of the Universit\`a di Trento during his research stays and support from the MIUR grant ``Dipartimenti di Eccellenza 2018-2022''.


\bibliographystyle{siam}       
\bibliography{BiblioFinal}   


\end{document}